\documentclass[journal]{IEEEtran}
\IEEEoverridecommandlockouts                              
\usepackage{graphics} 
\usepackage[pdftex]{graphicx}
\usepackage{adjustbox}
\usepackage{amsmath} 
\usepackage{amssymb}  
\usepackage{cite}
\usepackage{color}
\usepackage{float}
\usepackage{amsmath}
\usepackage{amssymb}
\usepackage{stackrel}
\usepackage{amsfonts}
\usepackage{MnSymbol}
\usepackage{multirow,hhline}
\usepackage{color}
\usepackage{algorithm}
\usepackage{algpseudocode}
\usepackage[table,xcdraw]{xcolor}
\def\BibTeX{{\rm B\kern-.05em{\sc i\kern-.025em b}\kern-.08em
    T\kern-.1667em\lower.7ex\hbox{E}\kern-.125emX}}
	
\begin{document}
		
		\title{A Review of Machine Learning-based Security in Cloud Computing}
		\author
		{
			Aptin~Babaei,~
			Parham~M.~Kebria,~\IEEEmembership{Member,~IEEE,}
			Mohsen~Moradi Dalvand,~
			and~Saeid~Nahavandi,~\IEEEmembership{Fellow,~IEEE}
			
			\thanks{A. Babaei, P. M. Kebria, and M. M. Dalvand are with the Institute for Intelligent Systems Research and Innovation (IISRI), Deakin University, Waurn Ponds, VIC 3216, Australia (email: a.babaei@deakin.edu.au; parham.kebria@deakin.edu.au; mohsen.m.dalvand@deakin.edu.au).}
			
			\thanks{S. Nahavandi is with Swinburne University of Technology, Hawthorn Melbourne, VIC 3122, Australia, also with the Harvard Paulson School of Engineering and Applied Sciences, Harvard University, Allston, MA 02134 USA (e-mail: saeid.nahavandi@ieee.org).}%
		}

	\maketitle
		
	\begin{abstract}\label{sec:Abstract}
		Cloud Computing (CC) is revolutionizing the way IT resources are delivered to users, allowing them to access and manage their systems with increased cost-effectiveness and simplified infrastructure. However, with the growth of CC comes a host of security risks, including threats to availability, integrity, and confidentiality.
		
		To address these challenges, Machine Learning (ML) is increasingly being used by Cloud Service Providers (CSPs) to reduce the need for human intervention in identifying and resolving security issues. With the ability to analyze vast amounts of data, and make high-accuracy predictions, ML can transform the way CSPs approach security.
		
		In this paper, we will explore some of the most recent research in the field of ML-based security in Cloud Computing. We will examine the features and effectiveness of a range of ML algorithms, highlighting their unique strengths and potential limitations. Our goal is to provide a comprehensive overview of the current state of ML in cloud security and to shed light on the exciting possibilities that this emerging field has to offer. 
	\end{abstract}
	

	\IEEEpeerreviewmaketitle

	\section{Introduction}\label{sec:Introduction}
		\IEEEPARstart{C}{}loud computing is a paradigm for delivering information technology services through the internet, rather than through a direct connection to a server. This delivery model allows on-demand access to computing resources, such as storage, networking, software, analytics, and intelligence, without needing physical infrastructure. Typically, customers are charged for cloud computing services based on usage, rather than a fixed rate, and can be classified into three Service Model categories: Infrastructure as a Service (IaaS), Platform as a Service (PaaS), and Software as a Service (SaaS).
		
		There are four deployment models for cloud computing: public, private, community, and hybrid. These models are comprised of various components, including providers, consumers, auditors, brokers, and carriers, each with distinct responsibilities\cite{chkirbene2019combined}.
		
		As cloud computing continues to gain popularity and expand, security concerns regarding the confidentiality, integrity, and availability of user data remain a pressing issue. To mitigate these concerns, it is imperative that cloud service providers adopt state-of-the-art technology and techniques to prevent data loss and unauthorized access while ensuring the availability of their services\cite{dwivedi2020study}.
		
		In this paper, we aim to conduct a comprehensive review of cloud computing service and deployment models, with a focus on security challenges and attacks. Furthermore, we will examine the role of machine learning in enhancing cloud security. Machine learning, a subfield of artificial intelligence, allows systems to learn and improve from experience without explicit programming. It involves the development of algorithms and models that can identify patterns in data and make predictions or decisions without human intervention.
		
		Our objective is to thoroughly examine the various machine learning techniques employed to detect, prevent, and resolve cloud security vulnerabilities. Despite numerous studies in this area, there is a lack of a comprehensive examination of the available machine learning algorithms in the context of cloud security. This paper will draw upon relevant literature and studies related to cloud computing and security, as well as the use of machine learning algorithms in cloud security.
		
		In section \ref{sec:Machine Learning in Cloud Security}, we will review different types of machine learning algorithms, including supervised, unsupervised, semi-supervised, and reinforcement algorithms, and discuss their potential benefits in enhancing cloud computing security. We will also examine some of the most commonly used machine learning algorithms and their features. Finally, in section \ref{sec:Conclusion}, we will provide a conclusion to our review and outline potential avenues for future research.
	
	\section{Cloud Computing}\label{sec:Cloud Computing}
		Cloud Computing has emerged as a transformative technology in the field of Information Technology (IT) over the past decade. It represents a paradigm shift in the way IT services are delivered and consumed by end-users. In Cloud Computing, a vast array of computing resources, including storage, computing power, and applications, are made available over the Internet on a pay-per-use basis. The computing resources are abstracted from the underlying physical infrastructure and made available to end-users through virtualized computing environments\cite{khorshed2011trust}.
		
		Cloud Computing is characterized by its scalability and on-demand availability. End-users can consume and scale computing resources according to their requirements without the need for investing in expensive physical infrastructure. This has the potential to offer significant cost savings for end-users, especially for small and medium-sized businesses. For example, online retailers can increase their computing resources during peak periods to meet the increased demand for their services, and then reduce them during periods of low demand, thus avoiding the costs associated with owning and maintaining physical infrastructure \cite{faheem2017Cloud}.
		
		From the perspective of Cloud Service Providers (CSPs), Cloud Computing is delivered through one of four deployment models: Public, Private, Hybrid, or Community Cloud. The deployment model chosen by a CSP depends on factors such as the level of control desired by the end-user, the level of trust required, and the level of customization required. Additionally, Cloud Computing can be further categorized into three main service models: Infrastructure as a Service (IaaS), Platform as a Service (PaaS), and Software as a Service (SaaS). In IaaS, end-users are responsible for managing their own virtual machines, operating systems, and applications. In contrast, in SaaS, the CSP is responsible for providing and maintaining the underlying infrastructure and software, leaving end-users with the minimum level of control over the computing environment\cite{kulkarni2012cloud}.
		
		While Cloud Computing offers many benefits, it also presents new security challenges. Ensuring the confidentiality, integrity, and availability of end-users' data are of paramount importance for CSPs. Attacks on Cloud Computing environments can have significant consequences, including data loss, unauthorized access, and disruption of service. To mitigate these risks, CSPs must implement robust security measures and technologies to protect their environment. One such technology is Machine Learning, which has the potential to significantly enhance the security of Cloud Computing. Machine Learning algorithms can automatically learn from data, identify patterns and anomalies, and make predictions or decisions without human intervention\cite{kulkarni2012cloud,zhangyandong2012cloud}.
		
		\subsection{Cloud Computing Service Models}\label{subsec:Cloud Computing Service Models}
\begin{figure}[t]
			\centering
			\includegraphics[width=\columnwidth]{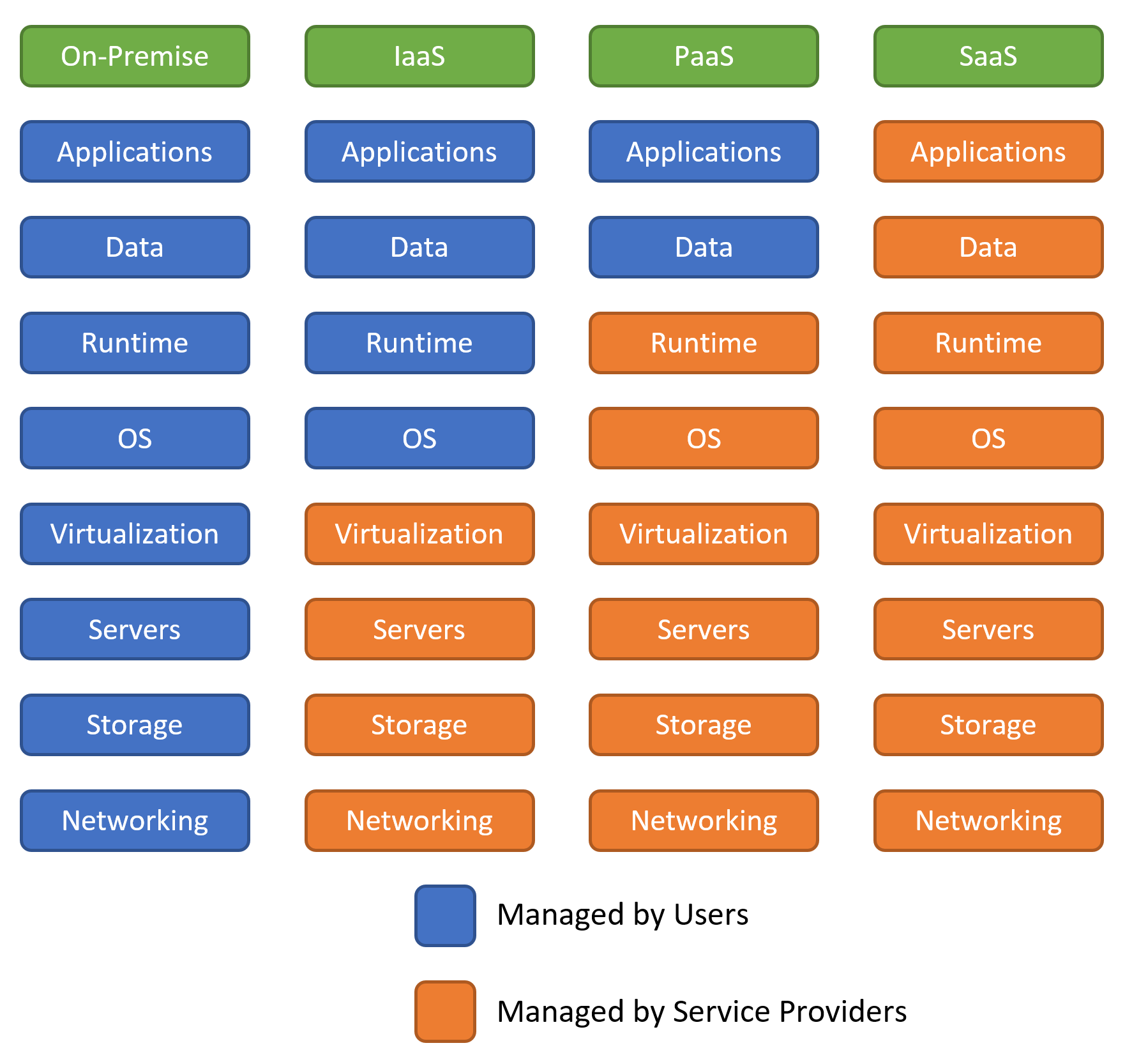}
			\caption{Different Cloud Computing Service Models and the level of responsibility for both parties}
			\label{fig:Cloud Computing Service Models}
		\end{figure}

			There are a few models that the Internet/Cloud providers are using to present their cloud services. Figure \ref{fig:Cloud Computing Service Models} shows IaaS, PaaS, and SaaS which are the most common models and represent the level of responsibilities of users and the service providers. Apart from these, there are more models like Content as a Service (CaaS), Database as a Service (DBaaS), Function as a service/Serverless (FaaS), etc\cite{hourani2018cloud}.
			
			\subsubsection{IaaS (Infrastructure as a Service)}\label{subsubsec:IaaS (Infrastructure as a Service)}
				The closest model to the on-site or on-premises IT infrastructure is IaaS. The cloud provider is responsible for the hardware infrastructure, including CPU, Storage, and Networking, and they should be taking care of related maintenance. The user would be responsible for installing the Operating System and all the applications related to their functions. This model has the highest level of responsibility for the users. Most of the providers of other models (PaaS, SaaS) use this model to sell their products to the end users.
				Microsoft and Amazon are some providers with this model in their portfolio and customers can use it. The providers use virtualization technologies to share their hardware with multiple customers \cite{zhangyandong2012cloud}.
			
			\subsubsection{PaaS (Platform as a Service)}\label{subsubsec:PaaS (Platform as a Service)}
				Customers who don't want to deal with hardware and operating system management, can use this model which the providers offer development tools, configuration management, and deployment platforms. Facebook development platform and Microsoft Windows platform are some examples of this model. Website developers who don't want to spend time installing software or OS can use this model \cite{kulkarni2012cloud}.
			
			\subsubsection{SaaS (Software as a Service)}\label{subsubsec:SaaS (Software as a Service)}
				The lowest level of responsibility for the users (mostly organizations or domestic users) can be found in this model. The users don't need to purchase software product licenses and they are not responsible for the maintenance or updating of them. Office 365 is an example of this model which provides hassle-free software to use on the web \cite{basu2018cloud}.
		
		\subsection{Cloud Computing Deployment Models}\label{subsec:Cloud Computing Deployment Models}
			From the service/cloud providers' perspective, the Cloud Computing has four different deployment models as followings:
			\subsubsection{Private Cloud}\label{subsubsec:Private Cloud}
				A private or internal cloud platform is designed specifically to be used by an organization and the services are not available to others. Enterprise companies or government departments with several branches are more suitable consumers for this model. This cloud model usually sits inside the enterprise or company itself which increases the security level and prevents external access but the investment cost is very high and not affordable for most consumers \cite{he2017machine}.
			\subsubsection{Public Cloud}\label{subsubsec:Public Cloud}
				A public or external cloud platform is available for all users rather than their use. It is usually built in multiple data centers for a greater level of redundancy. The users don't have any control over where their data gets saved. The service provider could use the geographically closer data center to save the users’ data to improve the latency of transmitting data \cite{garima2022machine}.
			\subsubsection{Hybrid}\label{subsubsec:Hybrid}
				A combination of private and public clouds with automation and orchestration facility, provides a hybrid Cloud Computing for consumers \cite{hourani2018cloud}. A hybrid cloud offers security and control like the private cloud also cost and elasticity similar to a public cloud. There are some concerns about data privacy and integrity while the data is traveling between the public and private sectors as they have different security levels and measurements.
			\subsubsection{Community}\label{subsubsec:Community}
			When a group of organizations has common interests like security requirements, policy, or compliance, they can use a community cloud model. In this model, the cloud is either owned by one of the organizations or a third party. In terms of the location, the community cloud infrastructure could be on-premise of one or multiple members of the community members or on a third-party's location \cite{basu2018cloud}.
			
			Table \ref{tbl:Cloud Computing Deployment Models} presents some advantages and disadvantages for each of the above models. \cite{butt2020review}.

			\begin{table}[b]
				\caption{Advantages and Disadvantages of Cloud Computing Deployment Models}
				\label{tbl:Cloud Computing Deployment Models}
				\centering
				\begin{adjustbox}{max width=\columnwidth}
					\begin{tabular}{|c|l|l|}
						\hline
						\multicolumn{1}{|l|}{Cloud   Models} & \multicolumn{1}{c|}{Advantages}        & \multicolumn{1}{c|}{Disadvantages}                           \\ \hline
						\multirow{3}{*}{Private}             & More reliability and control     & Lack of visibility and scalability                  \\
						& High security and privacy        & Limited Services                                    \\
						& Cost and energy efficient        & Security breach Vulnerabilities                     \\ \hline
						\multirow{3}{*}{Public}              & High scalability and flexibility & Less security of data \\
						& Cost-effective and reliable      & Potential for service outages                       \\
						& Location independence            &  Lower level of customizability                     \\ \hline
						\multirow{3}{*}{Hybrid}              & High scalability                 & Security compliance                                 \\
						& Low cost                         & Infrastructure dependent                            \\
						& More flexibility and security & More complex to set up and manage                   \\ \hline
						\multirow{3}{*}{Community}           & More secure than public Cloud    & Data segregation                                    \\
						& Lower cost than private Cloud    & Responsibilities allocation within the organization \\
						& More flexible and Scalable       &                                                     \\ \hline
					\end{tabular}
				\end{adjustbox}
			\end{table}
		
		\subsection{Cloud Computing Components}\label{subsec:Cloud Computing Components}
\begin{figure}[t]
				\centering
				\includegraphics[width=\columnwidth]{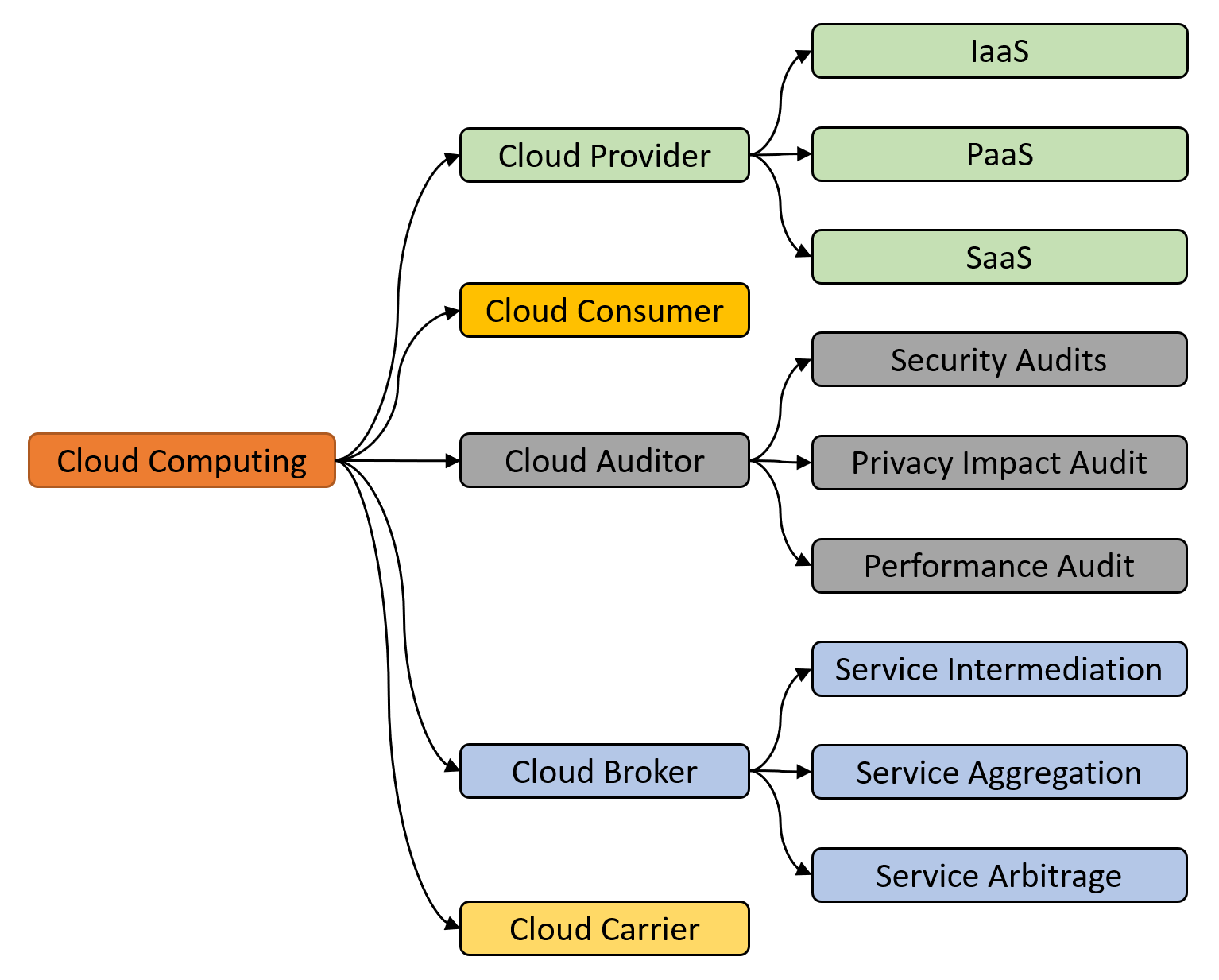}
				\caption{Cloud Computing Components}
				\label{fig:Cloud Computing Components}
			\end{figure}

			Cloud Computing as it is shown in figure \ref{fig:Cloud Computing Components} is a complex network including the following components \cite{butt2020review}:
			\begin{itemize}
				\item 
				Cloud Provider: This is an organization that creates a service that interested parties can access it. 
				\item 
				Cloud Consumer: This is the user of the Cloud Provider service.
				\item 
				Cloud Auditor: A party that is responsible to check the data framework, performance, and security of the cloud service.
				\item 
				Cloud Broker: An entity that manages the relationship between Cloud Provider and Consumer also deals with utilization, performance and delivery of service to the consumers.
				\item 
				Cloud Carrier: An infrastructure or medium which provides the network and transportation of cloud service to the Cloud Consumers from the Cloud Provider. 
			\end{itemize}
			
		\subsection{Cloud Computing Security Challenges}\label{subsec:Cloud Computing Security Challenges}
			Data security has always been one of the major topics in IT industry. Nowadays CSPs are responsible to provide the highest level of security for their consumer if they want to survive in this market. CSPs focus on three major factors to secure their cloud system: Confidentiality, Integrity and Availability (CIA)\cite{radwan2017cloud}. 
			\subsubsection{Confidentiality}\label{subsubsec:Confidentiality}
				Confidentiality is referring to the protection of data against unauthorized users. In a cloud environment, some users might try to access other users' data without permission. As the attacker could be on the same virtual machine as the victim user, the CSP is responsible to prevent unauthorized access from one tenant to another \cite{basu2018cloud}.
				CSP should provide a clear strategy for Data Storage, Deletion, Backup, Encryption, and Access privilege. If the stored data location is different from the user's location, in case of any attack, different laws would exist for each country. Also, if the deleted users' data, after canceling a subscription or account, get recovered by unauthorized users, the CSP is responsible for such un-trusted access\cite{sreelatha2020ensuring}.
			\subsubsection{Integrity}\label{subsubsec:Integrity}
				Integrity makes sure that the data has not been modified by an unauthorized user. Compared to confidentiality, integrity is focused on keeping the data intact from getting changed or modified by un-trusted users\cite{singh2017cloud}. 
				
				The attacker could use techniques like SQL injection to modify the database content or they can launch their malicious instance, instead of the user Virtual Machine (VM) instance on the cloud system. VM replication is another opportunity for the attacker which they can use to modify the data. If there are no proper security measures while the VM instance is getting replicated, the attacker can manipulate the process and modify the data during the replication. There are other VM attacks like rollback, escape, and hopping which are discussed in \cite{basu2018cloud}.
			\subsubsection{Availability}\label{subsubsec:Availability}
				This factor by far could be the most important security aspect for the service providers and their users. The CSPs need to make sure that the service is available close to 100\%. There are attacks like Denial of Service (DoS) that the attacker tries to make the service unavailable by sending rouge requests in bulk to occupy the servers’ CPU and memory.
				The cloud providers try to make their VMs available by using several data centers which are connected with high bandwidth links to each other. This way they can spread the servers and mitigate the security risks. Using techniques like "Virtual IP Address", "HoneyPot Zone" and "DDoS Mitigation" is helping the CSPs to overcome this challenge \cite{srikanth2021real}.
		\subsection{Cloud Computing Attacks}\label{subsec:Cloud Computing Attacks}
			There are a few vulnerabilities in Cloud Computing that could cause major threats which are virtualization, Unauthorized Access, Application Programming Interface (API), and browser vulnerabilities \cite{patel2013intrusion}. These vulnerabilities give the attacker a chance to launch an attack which could be categorized as followings based on the nature of the attack:
		
			\subsubsection{Denial of Service (DoS) attack}\label{subsubsec:Denial of Service (DoS) attack}
				As we discussed in the previous section, this is an attack to attempt to the availability of the service. As the source of this attack can be found, the attackers used Distributed Denial of Service to launch DoS attacks from multiple systems\cite{nassif2021machine}.
			\subsubsection{Zombie attack}\label{subsubsec:Zombie attack}
				In this scenario, the attacker uses an innocent user to storm the victim with bulk requests. This attack can affect the availability of the cloud system\cite{amara2017cloud}.
			\subsubsection{Man-In-The-Middle attack}\label{subsubsec:Man-In-The-Middle attack}
				As the name suggests, the attacker tries to get in the path between the users or between the user and the cloud system to manipulate the data or simply steal it\cite{aliyu2018detection}.
				
				Besides the above classification, we can categorize the attacks based on the cloud infrastructure section which the attackers aim for. In this case, the attacks can be categorized as \cite{garima2022machine}:
				\setcounter{subsubsection}{0}  
			\subsubsection{Network-based Attacks}\label{subsubsec:Network-based Attacks}
				The attacker aims to eavesdrop on the network traffic or try to restrict or alter it. Port Scanning, Spoofing, and Spamming are some examples in this category\cite{chiba2019intelligent}.
			\subsubsection{VM-based Attacks}\label{subsubsec:VM-based Attacks}
				Here the attackers try to manipulate the Virtual Machines' (VM) images. This can be done by injecting codes and/or changing the settings. VM Hyper Jacking and VM Escape are some examples of VM-based attacks\cite{lonea2013detecting}.
			\subsubsection{Storage-based Attacks}\label{subsubsec:Storage-based Attacks}
				Invading stored user information could be one of the most harmful attacks in which the attackers could get access to all the users’ data and then steal, change, or delete them. Data De-duplication, Data recovery, Data Scavenging, and Data Backup are some most known attacks in this category\cite{peng2012secure}.
			\subsubsection{Application-based Attacks}\label{subsubsec:Application-based Attacks}
				The application which is running on the cloud is vulnerable to attacks that can cause performance drops or information leakage. Malware infusion, Web Services, and Shared designs are some examples here\cite{sabahi2011virtualizationlevel}.
	\section{Machine Learning in Cloud Security}\label{sec:Machine Learning in Cloud Security}
		As Cloud Computing grows so fast, network management and controlling security are among the biggest concerns for the providers. In this circumstance automation especially using Machine Learning is a fast-growing technique to predict and prevent security risks and threats.
		
		Machine Learning (ML) is a sub-field of "Artificial Intelligence" responsible to fabricate underlying computational formulation and develop a statistical model based on the available dataset referred to as "Training Data". There are four main types of learning in ML which are Supervised, Unsupervised, Semi-supervised, and Reinforcement learning. 
		\subsection{Types of Machine Learning Algorithms}\label{subsec:Types of Machine Learning Algorithms}
			Based on the learning type, the Machine Learning algorithms can be categorized as the following which is illustrated in figure \ref{fig:Different types of Machine Learning algorithms} as well \cite{garima2022machine}:
			\begin{figure}[t]
				\centering
				\includegraphics[width=\columnwidth]{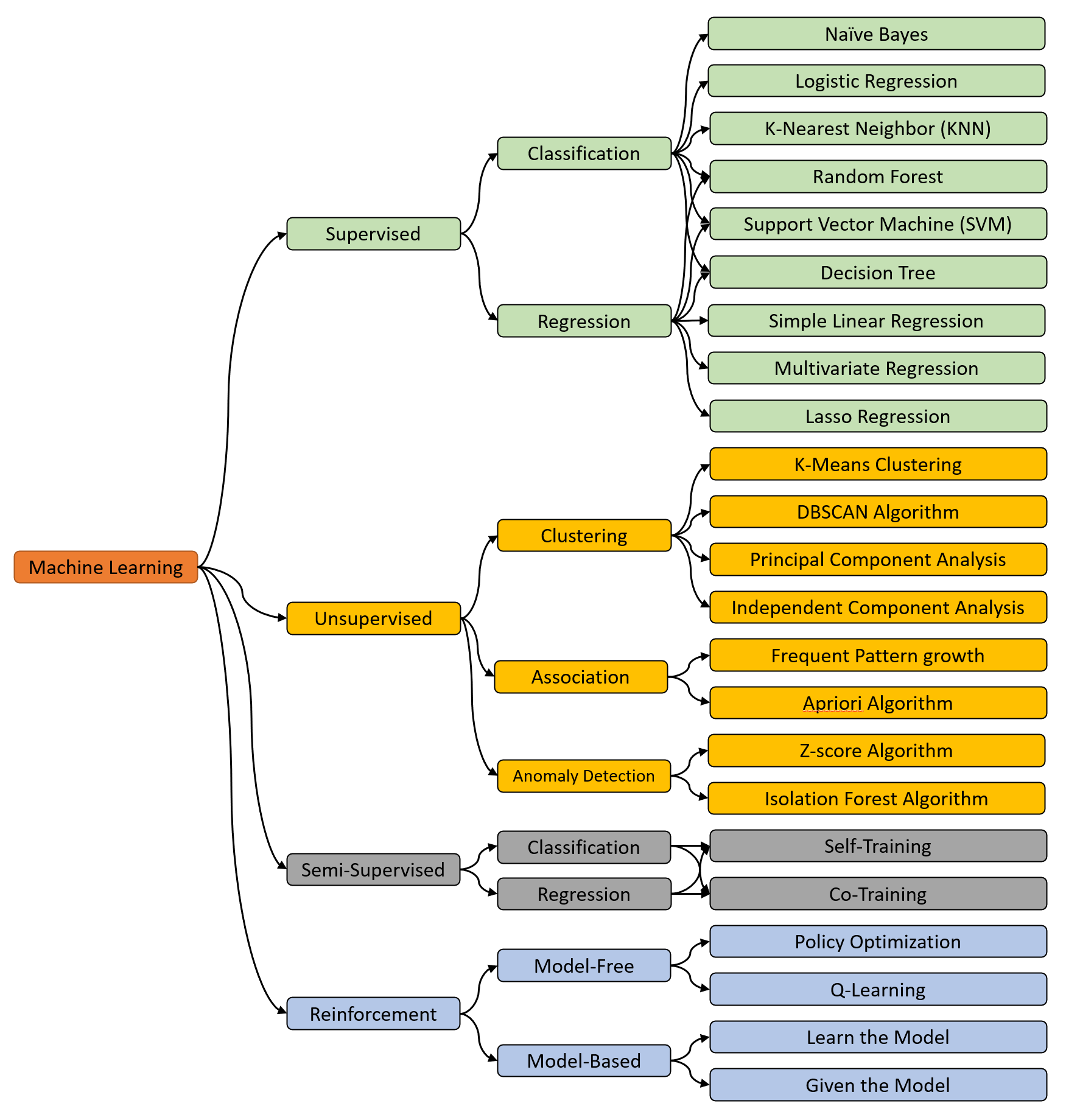}
				\caption{Different types of Machine Learning algorithms.}
				\label{fig:Different types of Machine Learning algorithms}
			\end{figure}
			\subsubsection{Supervised Learning}\label{subsubsec:Supervised Learning}
				In supervised learning, the system uses the provided training dataset as an instructor. This means the training dataset is already classified with an obvious and straightforward answer for each case or scenario. Thereafter the ML is given a new set of data to assess the outcome based on what it learned\cite{aboueata2019supervised}.
				This algorithm can train the model with labeled data for intrusion and/or anomalies over the cloud network besides the normal data. There are several common datasets available for this model like KDD, DRAPA, and UNSW.
				Based on the output, supervised learning can be categorized into the followings:
				\begin{itemize}
					\item 
					Classification: If the output data is categorical and from a certain list (True/False, Yes/No, Dog/Cat/Mouse) it is called a classification problem which could be a binary or multi-class problem. There are a few common examples for this category such as Naive Bayes, Decision Tree, Support Vector Machine (SVM), and Logistic Regression. According to \cite{subramanian2019focus,bhamare2016feasibility} Logistic Regression and Decision Tree are among the top three algorithms in terms of accuracy rate while the UNSW and ISOT datasets have been used.
					\item 
					Regression: If the problem output is continuous and numeric (e.g. a number between 0.0 to 100.0), the ML algorithm to use is called regression. There are algorithms like Linear Regression, Decision Tree, Lasso, and multivariate Regression in this category. In \cite{he2017machine}, it has been mentioned that the Linear Regression had a 94.36\% accuracy result compared to other algorithms used in the study\cite{bhamare2016feasibility}.
				\end{itemize}
	
			\subsubsection{Unsupervised Learning}\label{subsubsec:Unsupervised Learning}
				In this learning type, the system does not have any current labeled data to learn the pattern and prepare itself, instead, it has been designed to extract the pattern, classes, or categories from the data especially the data without any classification or labeling by itself \cite{wang2022cloud}.
				\begin{itemize}
					\item 
					Clustering: This algorithm tries to group similar data which has the same characteristics. Clustering is being used in image analysis, pattern recognition, and Machine Learning for its investigative data mining characteristic\cite{yasami2010novel}. 
					\item
					Anomaly Detection: In this algorithm, the system tries to identify instances that differ from the rest of the data. These differences could indicate potential issues or suggest interesting patterns be investigated\cite{dean2012ubl}.
					\item
					Association: The goal of this algorithm is to find the relationship/dependency between the data so that they can be grouped. The results can be used for businesses to make decisions about their marketing or product portfolio. For example, Amazon recommends related books or items to the users' search history\cite{bhatia2019unsupervised}.
				\end{itemize}
			\subsubsection{Semi-supervised Learning}\label{subsubsec:Semi-supervised Learning}
				As the name implies, in this model the system is not fully supervised so part of the training is with labeled data, and for the rest, the system should use unclassified or unlabeled data to find the patterns. This approach can be considered advantageous when procuring a sufficient amount of labeled data presents a challenge or incurs a significant expense. It needs to be observed how combining labeled and unlabeled data might change the learning behavior\cite{ravi2020semisupervisedlearningbased}.
				\begin{itemize}
					\item 
					Self-Training: In this procedure, we can take any method of supervised learning (Classification or Regression) and modify it to work as Semi-supervised Learning. First, we use some labeled data to train the model then we apply some labels to the remaining data and try to correct the labels through multiple iterations \cite{rosenberg2005semisupervised,xie2020selftraining}.
					\item 
					Co-Training: In this procedure, we create two classifiers based on two different views of the same data. Then each classifier labels the data and through different iterations, they help each other to improve the prediction accuracy. Finally, the two updated classifiers combine their results to create one classification output \cite{vale2022efficient,blum1998combining}.
				\end{itemize}
			\subsubsection{Reinforcement Learning}\label{subsubsec:Reinforcement Learning}
				Reinforcement learning is a sub-field of machine learning that deals with the problem of training autonomous agents to perform actions in an environment to optimize some notion of cumulative reward. This approach involves the agent continuously learning from its interactions with the environment to refine its decision-making policy\cite{qiang2011reinforcement}.
				
				The core idea behind reinforcement learning is to provide an agent with a scalar reward signal that it can use to learn an optimal behavior. The agent interacts with its environment by taking action, observing the resulting state, and receiving a reward. Over time, the agent uses this reward information to update its policy, which maps states to actions, to maximize the cumulative reward.
				
				Reinforcement learning has been successfully employed in a variety of real-world applications, including robotics, control systems, cloud security, and game-playing. It is particularly well-suited for problems in which the optimal solution is difficult to define, and the best course of action must be learned through trial and error and experience. \cite{garima2022machine}. 
				\begin{itemize}
					\item 
					Model-Based: This methodology involves constructing a model of the environment in which the agent operates, which can be utilized to simulate the outcomes of different actions. Model-based reinforcement learning can be applied to plan and estimate the expected reward of different sequences of actions before their execution\cite{li2020modelbased}.
					\item 
					Model-Free:	This approach focuses on directly learning a policy that maps states to actions without constructing a model of the environment. Examples of model-free reinforcement learning algorithms include Q-Learning and State-Action-Reward-State-Action(SARSA)\cite{kim2018eegbased}.
				\end{itemize}

				Table \ref{tbl:Example of some Machine Learning Algorithm with their Advantages and Disadvantages} is showing some examples of Machine Learning algorithms with their advantages and disadvantages.
	
				\begin{table*}[t]
					\caption{Example of some Machine Learning Algorithm with their Advantages and Disadvantages}
					\label{tbl:Example of some Machine Learning Algorithm with their Advantages and Disadvantages}
					\centering
					\begin{adjustbox}{max width=\textwidth}
						\begin{tabular}{|c|l|l|}
							\hline
							Algorithm                        & \multicolumn{1}{c|}{Adantages}                                           & \multicolumn{1}{c|}{Disadvantages}                                    \\ \hline
							\multirow{6}{*}{Random   Forest} & Effectively performs both classification and regression tasks            & Computationally intensive, especially with large datasets             \\
							& Handles high-dimensional data and is robust to outliers                  & Less interpretable than a single decision tree with many trees stored \\
							& Gives an estimate of feature importance for feature selection            & Prone to overfitting with noisy datasets                              \\
							& Less prone to overfitting than a single decision tree                    & Struggles with unbalanced datasets                                    \\
							& Can be used for both binary and multi-class classification problems      & Poor performance with fewer data points than features                 \\
							& Provides an estimate of prediction error for model selection             & Less effective with small feature sets                                \\ \hline
							\multirow{8}{*}{Decision   Tree} & Easy to understand and visualize                                         & Prone to overfitting with deep trees and limited data                 \\
							& Handles various types of data                                            & Unstable with slight changes in data resulting in different trees     \\
							& Requires minimal data preparation                                        & Not suitable for continuous variables, only binary splits             \\
							& Accurate handling of high-dimensional data                               & Biased towards high-level or frequently occurring features            \\
							& Used for both classification and regression tasks                        & Ineffective with noisy data and imbalanced datasets                   \\
							& Fast training and prediction times                                       & Less accurate compared to Random Forest or SVM                        \\
							& Easy to use with feature selection capability                            & Less interpretable than other algorithms                              \\
							& Can handle both binary and multi-class classification problems           &                                                                       \\ \hline
							\multirow{6}{*}{K-means}         & Simple implementation and understanding                                  & Assumes spherical and equally sized clusters                          \\
							& Efficient for large datasets                                             & Sensitive to initial cluster center points and scale of variables     \\
							& Handles diverse data types and distributions                             & Ineffective with categorical variables and noise/outliers             \\
							& Can be used for clustering, dimensionality reduction, and pre-processing & Number of clusters affects results                                    \\
							& Easy to interpret results and detect patterns                            & Not guaranteed to find global optimal solution                        \\
							& Versatile and applicable to a wide range of domains and tasks            & Not suitable for hierarchical, non-uniform, or non-convex clusters    \\ \hline
							\multirow{6}{*}{SVM}             & Performs classification, regression, and outlier detection tasks         & Sensitive to kernel choice and parameters.                            \\
							& Handles high-dimensional data effectively                                & Computationally intensive with large datasets.                        \\
							& Ideal for complex non-linear data boundary                               & No probabilistic estimates for classes.                               \\
							& Uses kernel functions for boundary optimization                          & Challenges with datasets with many features or instances.             \\
							& Memory efficient by using only a subset of training data                 & Poor performance with noisy data.                                     \\
							& Robust to overfitting with optimization avoiding overfitting             & Not effective for handling missing data.                              \\ \hline
						\end{tabular}
					\end{adjustbox}
				\end{table*}
	
		\subsection{Benefits of using Machine Learning in Cloud Security}\label{subsection:Machine Learning in Cloud Security}
			As has been mentioned in the previous sections, Cloud Computing is growing so fast and more consumers are joining this trend to use its benefits. Following this growth, Cloud Providers are under pressure to provide the required resources also maximize the security level for their users. In these circumstances, they would need some automation or new evolving technology to minimize human interaction, for faster action and less cost\cite{ntambu2021machine}. 
			Machine Learning has been used by most of the market leaders for cloud computing like AWS, Azure, IBM, and others. Below we can see some of the most important benefits of using Machine Learning in cloud security which have been illustrated briefly in figure \ref{fig:Machine Learning benefits in Cloud Security}:
			
			\begin{figure}[t]
				\centering
				\includegraphics[width=\columnwidth]{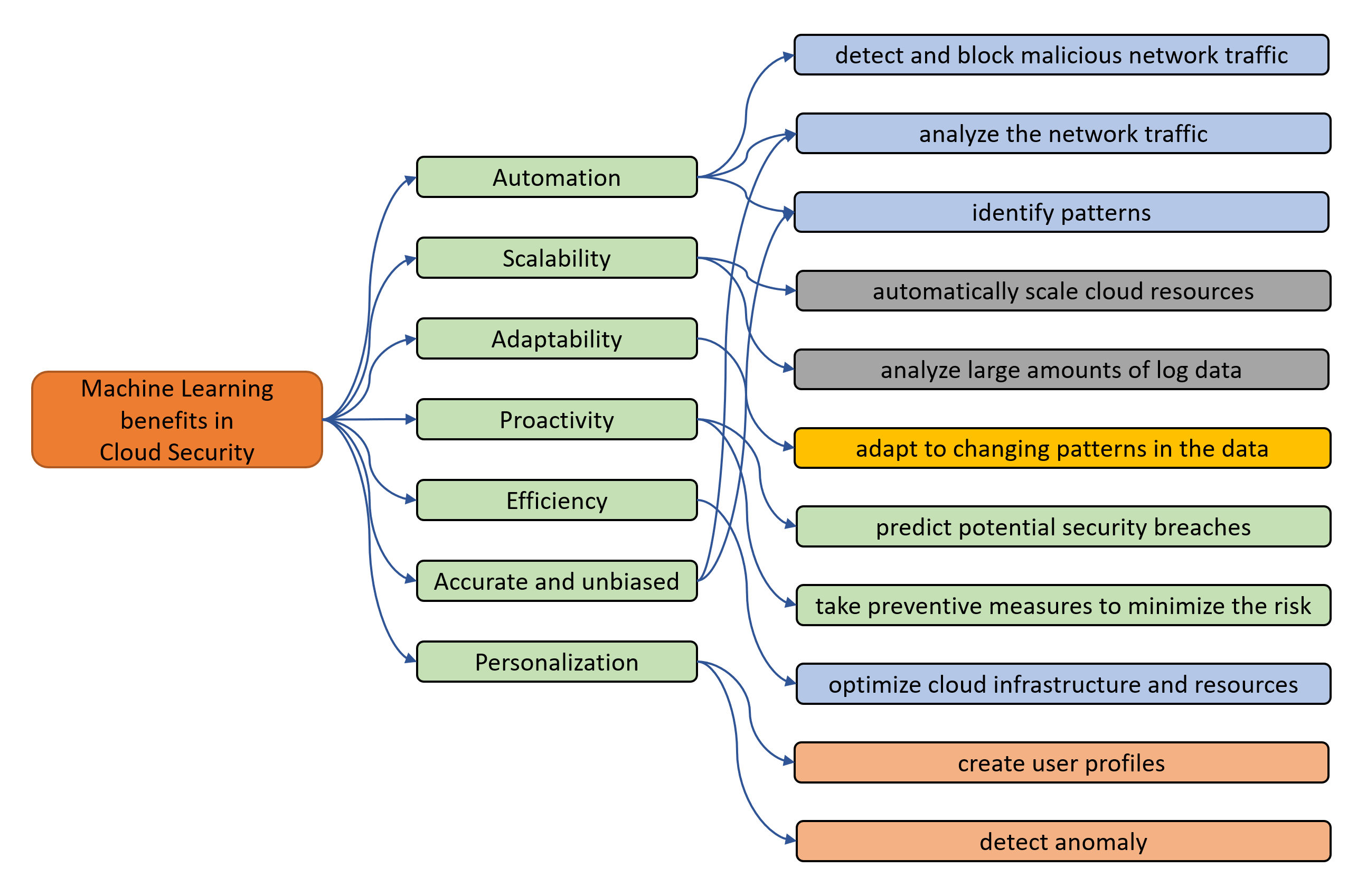}
				\caption{Machine Learning benefits in Cloud Security}
				\label{fig:Machine Learning benefits in Cloud Security}
			\end{figure}
	
			\subsubsection{Automation}\label{subsubsection:Automation}
				Using Machine Learning algorithms to automatically detect and block malicious network traffic, by analyzing the network traffic and identifying patterns that indicate a potential security threat. Also to automatically identify and respond to suspicious activity on cloud resources, by analyzing the usage patterns of the resources and identifying patterns that deviate from the norm. Once an unusual pattern is detected, the algorithm can automatically take appropriate action, such as revoking access or initiating an incident response. This can help organizations save time and resources by reducing the need for manual monitoring and intervention. Supervised learning algorithms such as Random Forest or Support Vector Machines to classify network traffic can be used here \cite{jain2022design}.
		
			\subsubsection{Scalability}\label{subsubsection:Scalability}
				Using Machine Learning algorithms to automatically scale cloud resources based on usage patterns, by analyzing the usage data from multiple sources, such as logs, network traffic, and system metrics, and identifying patterns that indicate when more resources are needed. Also to analyze large amounts of log data in real-time, by processing the log data and identifying patterns that indicate a potential security threat. Machine Learning algorithms such as Anomaly detection and Clustering can be used for this purpose \cite{salman2017machine}.
		
			\subsubsection{Adaptability}\label{subsubsection:Adaptability}
				Machine Learning algorithms can adapt to changing patterns in the data, making them more effective at detecting new and evolving threats. For example, Machine Learning algorithms can be trained on a dataset of known malware samples to detect new types of unknown malware or to identify new patterns of malicious behavior. This allows organizations to stay ahead of evolving threats and to respond quickly to new security risks. This can be done by using supervised learning algorithms such as Random Forest, Support Vector Machines, and Naive Bayes to classify files as malicious or benign \cite{nawrocki2019adaptable}.
		
			\subsubsection{Proactivity}\label{subsubsection:Proactivity}
				Using a supervised learning algorithm to predict potential security breaches, and to take preventive measures to minimize the risk of a successful attack. Also benefiting from a neural network to identify the patterns of unusual activity in cloud systems, such as login attempts from unusual locations or unexpected network traffic \cite{liu2018survey}.
		
			\subsubsection{Efficiency}\label{subsubsection:Efficiency}
				Machine Learning algorithms can be used to optimize cloud infrastructure and resources such as virtual machines, storage accounts, and network bandwidth; increasing the efficiency of the cloud environment and reducing costs. For example, Machine Learning algorithms such as linear regression, decision tree, and random forest can be used to automatically scale cloud resources based on usage patterns, or to identify and resolve inefficiencies in the cloud environment \cite{nassif2021machine}.
		
			\subsubsection{Accurate and unbiased}\label{subsubsection:Accurate and unbiased}
				Machine Learning algorithms can analyze large amounts of data to identify patterns that may be difficult for humans to detect, which can lead to more accurate security assessments. For example, a Machine Learning model can be trained to recognize patterns of network traffic that are commonly associated with a particular type of attack, such as a denial of service (DoS) attack. Once the model has been trained, it can then be used to analyze live network traffic and automatically block any traffic that matches the patterns associated with a known attack. Additionally, Machine Learning can improve the accuracy of security assessments by reducing the number of false positives generated by security systems. By analyzing data from multiple sources, Machine Learning models can more accurately identify security threats and reduce the number of false alarms\cite{bhamare2016feasibility}.
				
				Machine learning can also reduce the potential for bias in security decision-making by providing objective and data-driven insights into security-related activity. For example, Machine Learning can be used to analyze user behavior and identify patterns that indicate a potential security threat, such as a potential account compromise or insider threat, regardless of the user's identity, role, or privilege \cite{rawat2021secure}.
		
			\subsubsection{Personalization}\label{subsubsection:Personalization}
				Machine Learning algorithms can be used to create user profiles that capture the specific patterns of behavior of individual users or groups. This information can then be used to create more accurate and tailored security measures that are better suited to the needs of those users.
				Another way that Machine Learning can be used to provide personalization in cloud security is through the use of anomaly detection. Anomaly detection algorithms can be used to automatically identify and flag unusual or suspicious behavior, which can help to identify potential threats and respond to them more quickly and effectively.
				
				Machine Learning can also be used to create more sophisticated access controls that take into account the specific roles and permissions of different users. This can help to ensure that only authorized users have access to sensitive data and resources, while also helping to prevent unauthorized access \cite{alawneh2022personalized}.
	
		\subsection{Examples of Machine Learning Algorithms used in Cloud Security}
\begin{figure}[t]
			\centering
			\includegraphics[width=\columnwidth]{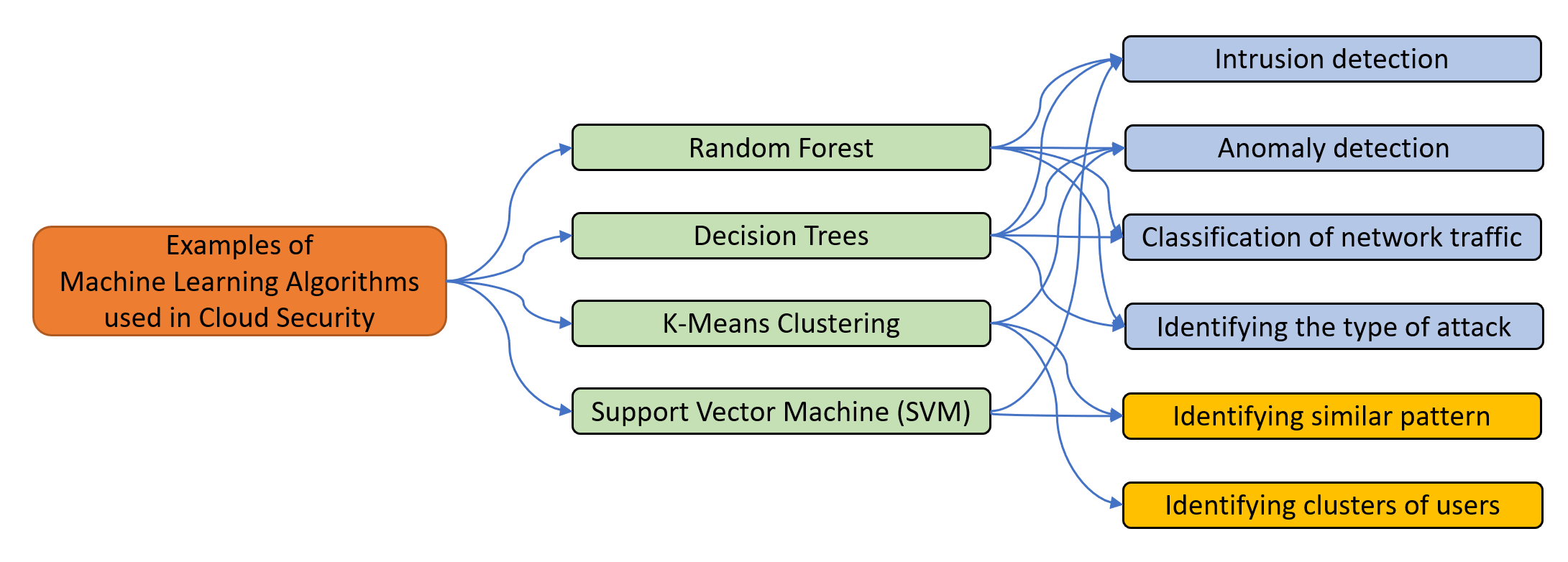}
			\caption{Examples of Machine Learning Algorithms used in Cloud Security}
			\label{fig:Examples of Machine Learning Algorithms used in Cloud Security}
		\end{figure}

			In this section, we are going to review some of the most common Machine Learning algorithms which have been used in Cloud Security. Besides the following examples, different vendors and organizations might use other algorithms or combine some of them to achieve their desired goals. Figure \ref{fig:Examples of Machine Learning Algorithms used in Cloud Security} shows these examples briefly.
			
			\subsubsection{Random Forest Algorithm}\label{subsubsection:Random Forest Algorithm}
				Random Forest is a Machine Learning algorithm that is used for both classification and regression tasks and it combines the predictions of multiple decision trees to make a final prediction. Each decision tree in the random forest is built using a different random subset of the training data, and the final prediction is made by taking a majority vote on the predictions of all the decision trees.
				
				Random Forest is a robust algorithm that can handle large datasets and high dimensional feature spaces. It is also able to handle missing or incomplete data, which is a common problem in cloud security. One of the key advantages of Random Forest is that it can identify the most important features in the dataset, which can be used to focus on the most critical areas of the network and improve the overall security of the system \cite{abdulkareem2021machine}.
		
				In terms of specific use cases, Random Forest can be used in Cloud Security for a variety of tasks, such as:
				\begin{itemize}
					\item 
					Intrusion detection: Random Forest can be trained on a dataset of normal network activity, and then used to identify patterns of activity that deviate from the normal behavior. These patterns can indicate the presence of an intrusion attempt, and the algorithm can alert the security team to take action.
					\item 
					Anomaly detection: Random Forest can be used to identify unusual or unusual patterns in data that might indicate a security threat. This can include identifying unusual behavior in network traffic, such as a sudden increase in traffic from a specific IP address, or identifying unexpected changes in the configuration of a system.
					\item 
					Classification of network traffic: Random Forest can be used to classify network traffic as normal or malicious, based on features such as source IP address, destination IP address, port number, and packet size. This can help to quickly identify and respond to potential security threats.
					\item 
					Identifying the type of attack: Random Forest can also be used to classify the type of attack, whether it's a DDoS, malware, or phishing attack.
					The Random Forest algorithm is also a powerful tool for feature selection, which is the process of identifying the most important features in a dataset. It can be used to identify the most important features in network traffic data, such as specific IP addresses or ports, that are most likely to indicate a security threat. This can help to focus security efforts on the most critical areas of the network and improve the overall security of the system.
				\end{itemize}
		
			\subsubsection{Decision Trees Algorithm}\label{subsubsection:Decision Trees}
				Decision Trees are simple and intuitive algorithms that can be easily understood and interpreted by humans. The tree-like structure of the algorithm makes it easy to visualize the decision-making process, and the branches of the tree represent the different decisions that are made based on the input features. This makes the algorithm well-suited for applications where transparency is important, such as cloud security \cite{subramanian2019focus}.
				
				A Decision Tree is used for both classification and regression tasks in Machine Learning. It is a type of supervised learning algorithm that can be used to make predictions based on input data. The decision tree algorithm works by recursively partitioning the data into smaller subsets based on the values of the input features. At each step, the algorithm selects the feature that best separates the data into the target classes. The process continues until a stopping criterion is met, such as a maximum tree depth or a minimum number of samples in a leaf node. One of the key advantages of Decision Trees is that they can handle both continuous and categorical data, making them versatile and suitable for a wide range of applications. They can also handle missing data, a common cloud security problem.
				Decision Trees Algorithm can be used in similar scenarios as the Random Forest Algorithm, such as Intrusion and Anomaly detection, Classification of network traffic, and Identifying the type of attack and suspicious behavior\cite{fitni2020implementation}.
			
			\subsubsection{K-Means Clustering Algorithm}\label{subsubsection:K-Means Clustering}
				K-Means Clustering is an unsupervised Machine Learning algorithm used for clustering data into groups or clusters which is widely used in Machine Learning and data mining \cite{win2018big}. Some more details about the algorithm include:
				\begin{itemize}
					\item 
					Centroid-based: K-Means Clustering is a centroid-based algorithm, which means that it works by defining clusters based on the mean of the points in the cluster. The algorithm initializes k cluster centroids, where k is the number of clusters that you want to create, and then assigns each data point to the nearest centroid.
					\item 
					Iterative: The algorithm is iterative, meaning that it repeatedly updates the cluster centroids and reassigns data points to clusters until the centroids no longer change or a maximum number of iterations is reached.
					\item 
					Determining k: One of the challenges in using K-Means Clustering is determining the optimal number of clusters (k) for a given dataset. Several methods can be used to determine the optimal number of clusters, such as the elbow method, silhouette analysis, and gap statistics.
					\item 
					Assumptions: K-Means Clustering makes some assumptions about the data, such as the clusters being spherical and having roughly the same size and density. These assumptions may not hold true for all datasets, and in such cases, other clustering algorithms such as hierarchical clustering or density-based clustering may be more appropriate.
					\item 
					Computationally efficient: K-Means Clustering is a computationally efficient algorithm, making it suitable for large datasets. However, storing the cluster centroids and data points requires a large amount of memory.
					\item
					Limitations: K-Means Clustering is sensitive to the initial placement of the centroids, and can lead to suboptimal solutions if the initial centroids are not chosen carefully. Additionally, it is not well suited for datasets with non-numeric variables or categorical data, and it also assumes that the clusters have similar sizes and densities, which may not be the case for all datasets.
				\end{itemize}
				K-Means Clustering is a widely used algorithm for clustering data into groups or clusters. It is computationally efficient, making it well-suited for large datasets. However, it has some limitations and assumptions that should be considered when using it, such as the difficulty of determining the optimal number of clusters and the sensitivity to the initial placement of the centroids\cite{muniyandi2012network}.
				
				In terms of use cases, K-Means Clustering can be used in cloud security for a variety of tasks, such as Anomaly detection, Identifying similar patterns for grouping similar malicious IP addresses, or for grouping similar types of attacks, also Identifying clusters of users based on their behavior such as a user accessing sensitive data at odd hours or from an unusual location.
		
			\subsubsection{Support Vector Machine (SVM) Algorithm}\label{subsubsection:SVM}
				Support Vector Machines (SVMs) are a type of supervised learning algorithm that can be used for classification and regression tasks. They work by mapping the input data into a high-dimensional feature space, where a hyperplane can be used to separate the different classes. The key idea is to find the hyperplane that maximizes the margin, which is the distance between the hyperplane and the closest data points from each class, also known as support vectors. These support vectors define the decision boundary, and any new data point can be classified based on which side of the boundary it falls on \cite{nassif2021machine}.
				One of the main advantages of SVMs is that they can handle non-linearly separable data by transforming the input data into a higher dimensional space using a technique called the kernel trick. This allows the algorithm to find a linear boundary in the transformed space that separates the data.
				
				In cloud security, SVMs can be used for Intrusion Detection by analyzing network traffic and identifying patterns that indicate malicious activity. The algorithm can be trained on labeled data, such as normal network traffic and known intrusion attempts, to learn the characteristics of each class. After training, it can classify new, unseen network packets as normal or abnormal. This can help to detect and prevent attacks on the cloud infrastructure, such as denial of service (DoS) attacks, malware infections, and unauthorized access attempts\cite{wang2022cloud}.
				
				Additionally, SVMs can be used to classify and detect malicious files, such as malware, in the cloud. By analyzing the content of the files and extracting features, the algorithm can be trained on labeled data to learn the characteristics of benign and malicious files.
	
	\section{Conclusion}\label{sec:Conclusion}
	In this paper, we have reviewed Cloud Computing technology and talked about different Service and Deployment Models. Then we reviewed the Security Challenges and common attacks in the Cloud Computing environment.		
	After that, the focus was on Machine Learning and we reviewed the types of Machine Learning Algorithms, and the benefits of using ML in Cloud Security, finally, we discussed the most common ML algorithms used in Cloud Security.
	
	Besides the points that we covered, there are more use cases like Cloud-based security orchestration and automation (SAO), assessment, incident, compliance and/or configuration management, Cloud Security Analytics, Governance, Monitoring, Cloud Access Security Brokers (CASB), and others which we are going to focus on them in future works.
	Also, we are going to dive deep into some of the ML algorithms and try to implement some scenarios with publicly available datasets.
	
	\bibliography{C:/Users/bc209/OneDrive/Study/PHD/Deakin/Bibliography.bib}
	\bibliographystyle{IEEEtran}
	
\end{document}